\documentclass{article}

\usepackage{amssymb}
\usepackage{graphicx}
\usepackage{amsmath}
\usepackage{amsfonts}
\usepackage{psfrag}

\psfrag{i}{$i$}
\psfrag{j}{$(ij)$}
\psfrag{a}{$a$}
\psfrag{b}{$b$}
\psfrag{g}{\mbox{$g_{ab}$}}
\psfrag{h}{$g_{ba}$}
\psfrag{m}{$g_{ij}$}
\psfrag{k}{$g_{ik}$}
\psfrag{l}{$g_{aij}$}
\psfrag{n}{$g_{aik}$}
\psfrag{p}{$(ik)$}
\psfrag{q}{$(ij)$}
\psfrag{r}{$e_{a(ij)}$}
\psfrag{s}{$e_{ij}$}

\newcommand{\Ref}[1]{(\ref{#1})}

\newcommand{\eqa}{\begin{eqnarray}}
\newcommand{\neqa}{\end{eqnarray}}

\newcommand{\no}{\nonumber\\}
\renewcommand{\alpha}{{a}}
\renewcommand{\beta}{{b}}
\renewcommand{\gamma}{{c}}

\begin{document}

\title{\LARGE\bf{Spinfoam 2d quantum gravity and discrete
bundles}} \author{ {\normalsize{Daniele Oriti${}^{a}$, Carlo
Rovelli${}^{bc}$, Simone Speziale${}^{ac}$} } \\[1mm] \em\small{${}^a$
DAMTP, Centre for Mathematical Sciences, University of Cambridge CB3
0WA, Cambridge, EU.}\\[-1mm] \em\small{${}^b$ Centre de Physique
Th\'eorique de Luminy, Universit\'e de la M\'editerran\'ee, F-13288
Marseille, EU. }\\[-1mm] \em \small{${}^c$ Dipartimento di Fisica
dell'Universit\`a ``La Sapienza", INFN Sez.\,Roma1, I-00185 Roma,
EU.}} \date{\small\today} \maketitle

\begin{abstract}

In 4 dimensions, general relativity can be formulated as a constrained
$BF$ theory; we show that the same is true in 2 dimensions.  We
describe a spinfoam quantization of this constrained $BF$-formulation
of 2d riemannian general relativity, obtained using the Barrett-Crane
technique of imposing the constraint as a restriction on the
representations summed over.  We obtain the expected partition
function, thus providing support for the viability of the technique.
The result requires the nontrivial topology of the bundle where the
gravitational connection is defined, to be taken into account.  For
this purpose, we study the definition of a principal bundle over a
simplicial base space.  The model sheds light also on several other
features of spinfoam quantum gravity: the reality of the partition
function, the geometrical interpretation of the Newton constant, and
the issue of possible finiteness of the partition function of
quantum general relativity.

\end{abstract}

\section{Introduction}

Spinfoam formulations of quantum gravity can be derived in several
ways (for reviews see \cite{Daniele1,Perez2,book}.)  One technique is
based on the relation between general relativity (GR) and $BF$ theory:
GR can be expressed as a $BF$ theory plus a constraint, called the
``Plebanski constraint".  $BF$ theory has a well-understood covariant
quantization, given by a Feynman-like sum over representations of the
gauge group, associated with (faces of a two-complex interpreted as
the two-skeleton of) a simplicial discretization of the spacetime
manifold.  A spinfoam model for quantum GR can be obtained enforcing
the Plebanski constraint into the spinfoam sum and, eventually,
summing over two-complexes.  The constraint can be enforced as a
restriction on the class of representations summed over.  This way of
enforcing the constraint is called the Barrett-Crane technique and the
resulting model is called the Barrett-Crane model (although Barrett
and Crane's original rationale for restricting the class of
representations was different \cite{Barrett1}.)  Variants of this
model have been studied, both in the riemannian and the lorentzian
context; some of these have remarkable finiteness properties
\cite{PR,finiteness,finiteness2} and, although many issues remain
unclear, the spinfoam formalism is often regarded as the covariant
version of loop quantum gravity \cite{book,ThiemannThesis} and as
candidate for computing transition amplitudes in quantum gravity.

A source of concern about this approach is the legitimacy of the
Barrett-Crane technique.  Ideally, one would like to test it by
analyzing the physics predicted by the model, but our understanding of
background independent physics is still insufficient for this.  In
this situation, it is interesting to test whether the technique works
in a context which is simple and under control.  This is what we do in
this paper.

Indeed, we show below that in two dimensions (2d), GR admits a
formulation as a constrained $BF$ theory, analogous to the 4d theory.
Riemannian 2d quantum GR has been studied with several techniques (see
\cite{ambjorn} and references therein) and its partition function is
known.  Quantum $BF$ theory in 2d has been studied \cite{Blau}; its
quantization using the spinfoam formalism and group field theory (GFT)
techniques \cite{GFT,mikecarlogft} was considered in \cite{carlo}.  It
is therefore natural to study the spinfoam quantization of 2d
riemannian GR using the Barrett-Crane technique, in order to test such
a technique.  Here we do so and we find that the Barrett-Crane
technique works well in the 2d context, and it yields the expected
partition function.  Remarkably, the partition function of 2d quantum
GR turns out to be finite and well-defined, in spite of the fact
that the $BF$ partition function diverges; thus, the Barrett-Crane
technique improves the finiteness of the spinfoam sum,
in 2d. We will comment on the interpretation of this fact
and on its relevance for the 4d case.

In spite of the simplicity of the 2d model, the way the quantization
works is rather nontrivial.  In order for the spinfoam model to give
the correct partition function, it is necessary to take the nontrivial
topology of the principal bundle on which the $BF$ theory is defined
(which is distinct from the topology of the base manifold, namely the
topology of spacetime) into account.  A naive quantization that
disregards the bundle topology gives an incorrect result.
This fact raises the question of whether similar aspects
should be considered also in higher dimensions.

Now, spinfoam quantization requires a discretization of the spacetime
manifold.  The definition of a nontrivial principal bundle over a
discrete base is not immediate.  In the continuum, the nontriviality
of the bundle is encoded in {\em continuous} transition functions on
the overlaps between open sets covering the base; in a discrete
principal bundle, suitable {\em discrete} data must encode the bundle
topology.  We thus need to define the geometry of such a discrete
principal bundle, in order to construct the model.  We do this in
Section \ref{math} and, in a different and more detailed manner, in
the Appendix.

A second issue we need to address is the scaling of the connection,
namely the ``size" of the internal gauge group.  It is convenient to
distinguish between the group elements that define the spinfoam model
and the parallel propagators of the gravitational connection.  The
scaling factor between the two is where the Planck constant and the
Newton constant are hidden, in a spinfoam model.

With all this technology at hand, the quantization of $BF$ theory is
straightforward, and the Barrett-Crane technique yields the expected
2d gravity partition function, with no divergences appearing.  The
spinfoam model naturally yields the real partition function, related
to the cosinus of the action instead than its complex exponential,
precisely as it happens for the 3d Ponzano-Regge model
\cite{Ponzano:1968}.  The reason for this and the relation with the
orientation of the manifold and time-inversion, are transparent in 2d.
Thus the 2d case sheds light and provides support for the spinfoam
approach to quantum gravity.

In the next section we give a short review of 2d gravity, we
illustrate its relation with $BF$ theory and we point out the
relevance of the bundle topology.  In Section \ref{torus} we review
the spinfoam quantization of $BF$ theory and we present the spinfoam
quantization of 2d GR in the case in which spacetime has the topology
of the torus and the bundle is trivial.  Section \ref{math} develops
the mathematical machinery to deal with nontrivial discretized
bundles.  In Section \ref{infine} we apply this machinery to 2d GR for
a general topology, and we show that the Barrett-Crane technique
yields a finite partition function, with the expected form.
In section \ref{conclusion} we summarize our results, and we
discuss the relevance of the issues here raised for the higher dimensional
cases.   In the Appendix, we develop
a definition of discrete bundles which may be more appropriate to deal
with higher dimensions and with the GFT formalism, and we suggest a
possibility to include the data characterizing the bundle topology
into the GFT formalism, by modifying the kinetic term of the GFT
action.

\section{General Relativity as a constrained $BF$ theory in 2d}
\label{GRcomeBFsection}

\subsection{2d GR}

We consider riemannian general relativity in two dimensions.  The
theory is defined by the Einstein-Hilbert action
\begin{equation}
 \label{EH} S_{\rm GR}[g]=\frac{1}{16\pi G}\int d^2x \,\sqrt{\det g}\
R
\end{equation}
for a 2d riemannian metric field $g_{\mu\nu}(x),\
\mu,\nu=1,2$.  Here $G$ is the Newton constant, $R$ the Ricci scalar
and $\det g$ the determinant of $g_{\mu\nu}$.  In 2d, the
Riemann tensor has only one independent component, say $R_{1212}$; the
Ricci tensor and the the Ricci scalar are determined by
\begin{equation}
(\det g)\, R_{\mu\nu}=g_{\mu\nu}{R_{1212}}, \hspace{1cm}
(\det g)\, R= 2\,R_{1212}, \end{equation} and therefore the Einstein
tensor vanishes identically
\begin{equation}
 R_{\mu\nu}-\frac{1}{2}
 g_{\mu\nu}R\equiv 0.
\end{equation}
It follows that the vacuum Einstein equations are satisfied by {\em
any} metric.  In fact, the action (\ref{EH}) is invariant under local
variations of the metric.  In particular, on a compact 2d surface $M$,
the Gauss-Bonnet theorem (see for instance \cite{chern1}) states that
\begin{equation}
\label{EH1}
\int_Md^2x\,\sqrt{\det g}\,R= 4\pi\chi(M),
\end{equation}
where $\chi(M)$ is the Euler's characteristic of $M$, an integer that
depends only on the genus $g$ of the manifold, namely the number of
holes, via $\chi=2-2g$. Hence on a compact manifold $M$
\begin{equation}
\label{Schi}
S_{\rm GR}[g] = \frac{\chi(M)}{4G}.
\end{equation}

Formally, the theory can be covariantly quantized by writing the
partition function
\begin{equation}
\label{Zgenrel}
\tilde{Z}_{\rm GR}[M]=\int {\cal D}g\ e^{\frac{i}{\hbar}S_{\rm GR}[g]},
\end{equation}
where ${\cal D}g$ is an unknown measure on the infinite dimensional
space of metrics modulo diffeomorphisms at fixed topology. We may
assume that this measure is normalized, that is
\begin{equation}
\int {\cal D}g = 1.
\label{norm}
\end{equation}
Notice that the gravitational constant $G$ in 2d has the inverse
physical dimensions of an action, and that the product $\hbar G$ is
dimensionless.  Since the action does not depend on the particular
choice of the metric, but only on the topology of the manifold,
\Ref{Schi} gives
\begin{equation}
\tilde{Z}_{\rm GR}[M] =\int {\cal D}g\,e^{\frac{i\chi
(M)}{4\hbar G}}= e^{\frac{i\chi (M)}{4\hbar G}}\int {\cal D}g.
\end{equation}
Using the assumption (\ref{norm}) that ${\cal D}g$ is normalized, we
obtain the following partition function for 2d riemannian general
relativity
\begin{equation}
\label{PartGR}
\tilde{Z}_{\rm GR}[M]=e^{\frac{i\chi (M)}{4\hbar G}}.
\end{equation}

In a background dependent quantum theory, the sign of the exponent in
(\ref{Zgenrel}) is determined by time orientation: it is reversed
under $t\to-t$.  The partition function and all transition amplitudes
go into their complex conjugates under time reversal.  The situation
in a background independent theory is less clear.  It is often argued
that both directions of time evolution, or, equivalently, both
orientations of the manifold $M$, must be summed over in the Feynman
integral.  This yields a real partition function, and real transition
amplitudes, and is consistent with the fact that the Wheeler-DeWitt
equation has real solutions.  See \cite{book} and \cite{LO} for a more
detailed discussion.  Here, we consider also the real partition
function
\begin{equation}
\label{Zgenrelreal}
Z_{\rm GR}[M]\ =\ \int {\cal D}g\,\left(e^{\frac{i}{\hbar}S_{\rm GR}[g]} +
e^{-\frac{i}{\hbar}S_{\rm GR}[g]}\right)\ = \ 2\
\cos\left(\frac{\chi(M)}{4\hbar G}\right),
\end{equation}
which is (twice) the real part of (\ref{PartGR}).  Our task is to
recover this result from a spinfoam quantization of the theory,
without using an ill-defined infinite-dimensional measure, and without
encountering any divergence.

We use the ``dyad" formalism, namely the 2d analog to the tetrad
formalism.  Let $e^i(x)=e_{\mu}^i(x)dx^\mu,\ i=1,2$, be a dyad
one-form field (with $\det e \ne 0$) related to the metric by
$g_{\mu\nu}(x)=e_{\mu}^i(x)e_{\nu}^j(x)\delta_{ij}$.  Let
$\omega^i{}_{j}(x)= \omega_{\mu}^i{}_{j}(x) dx^\mu$ be the
corresponding spin-connection one-form, defined by the Cartan
structure equation
\begin{equation}
de^i+\omega^i{}_{j}\wedge e^j=0.
\label{cartan}
\end{equation}
$\omega^i{}_{j}$ is an abelian $SO(2)$ connection on the frame bundle.
Its curvature is the Lie-algebra valued two-form $F^i{}_j
=\frac{1}{2}F^i{ }_{j\mu\nu} dx^\mu \wedge dx^\nu$ defined by
$F^i{}_{j}=d\omega^i{}_{j}$.  Using the generator $\tau$ of the $so(2)$
algebra, with components
\begin{equation}
\tau^i{}_{j}={\scriptscriptstyle\left(\begin{array}{cc} 0 &
1 \\ -1 & 0 \end{array}\right)}\ ,
\label{tau}
\end{equation}
we can write $ \omega^i{}_{j}= \omega \tau^i{}_{j} $ where $\omega$ is
a real valued one-form; and $ F^i{}_{j}=f\tau^i{}_{j} $ where
$f=d\omega$ is a real valued two-form.  We shall denote $so(2)$
matrices with capital letters, thus $F=f\tau$ has
matrix components $f\tau^i{}_{j}$, and so on.

The curvature of the spin connection, $F$, is related to the Riemann
curvature by $ e^i_\mu R^\mu{ }_{\nu\rho\sigma}= e^j_{\nu} F^i{
}_{j\rho\sigma}$.  Using this, it is easy to see that in terms of
these fields the action (\ref{EH}) reads
\begin{equation}
\label{EHeo} S_{\rm GR}[e]\ =\ \frac{1}{8\pi G}\int {\rm sgn}(e) \ f
\ =\ \frac{{\rm sgn}(e)}{8\pi G}  \int  d\omega\ ,
\end{equation}
where ${\rm sgn}(e)$ is the sign of the determinant of $e_{\mu}^i$.
Because of Stokes theorem, this action over an open set is a boundary
term and is therefore invariant under local field variations.  Hence
{\em any} dyad field, or {\em any} connection $\omega$ solves the
equations of motion locally.

Notice that if we formally define the quantum theory in terms in this
formalism
\begin{equation}
\label{Zgenrele}
Z_{\rm GR}[M]=\int {\cal D}e\ e^{\frac{i}{\hbar}S_{\rm GR}[e]}\;,
\end{equation}
we must sum over both signs of $\det e$ and therefore we obtain the
real partition function (\ref{Zgenrelreal}).  Clearly, a dyad with
negative determinant, such as $e^1=-dx^1, e^2=dx^2$, is equivalent to
a ``time-reversed`" field, or an opposite orientation of $M$.
Therefore the dyad formalism leads naturally to the real partiction
function.  This can be compared with the case the Ponzano-Regge
theory, where the same phenomenon happens.

\subsection{Constrained $BF$ theory}

We now show that 2d riemannian GR can be expressed as a constrained
$BF$, like 4d GR.  Let $\omega$ be a $SO(2)$
connection and $B=b\tau$ a scalar field with values in the
$so(2)$ algebra.  We consider the $BF$ theory defined by the action
\begin{equation}
\label{BF}
S_{BF}[\omega,B]
= -\frac{1}{2}k \int {\rm tr }\,[BF]
= k\int bf
\end{equation}
where $f=d\omega$.  Here $k$ a constant with the physical dimensions
of an action.\footnotemark\footnotetext{Using dimensionless
coordinates, the covariant derivative, the connection and the
curvature are dimensionless.} The factor -1/2 comes from ${\rm
tr}[\tau\tau]=-2$.  The equations of motion are the standard $BF$
equations
\begin{equation}
\label{em1}
dB(x)=0,
\end{equation}
and
\begin{equation}
\label{em2}
F(x)=0,
\end{equation}
which are solved by any flat $\omega$ and constant $B$.  The theory is
topological, like 2d GR, namely it has no local degrees of freedom,
but is different from 2d GR. In fact, we can identify the connection
$\omega$ of the $BF$ theory with the spin connection $\omega$ of GR,
but while the equations of motion of 2d GR are solved by {\em any}
$\omega$, the equations of motion of $BF$ are solved only by {\em
flat} connections.  Hence GR and $BF$ are distinct theories in 2d, as
they are distinct theories in 4d.

Recall that in 4d GR can be obtained from $BF$ theory by adding a
constraint.  Schematically, the action of 4d GR can be written as
\begin{equation}
S_{\rm GR}[e,\omega]=\frac{1}{16\pi G} \int  {\rm tr} [*(e\wedge
e)\wedge F]
\end{equation}
where $e$ is the (Minkowski valued) tetrad (one-form) field, $\omega$
is the corresponding spin connection, $F$ its curvature and the star
indicates the Hodge dual.  We can add a constraint to a $SO(4)$ $BF$
action, as follows
\begin{equation}
S_{\rm GR}[B,\omega,\lambda]=k\int {\rm tr}\, [B\wedge F - \lambda
B\wedge B].
\end{equation}
Varying the Lagrange multiplier $\lambda$ we obtain a Plebanski
constraint of the form $B\wedge B=0$, which is solved when $B$ has the
form $B=*(e\wedge e)$.  Inserting this back into action, and
identifying $k$ with $1/16\pi G$ gives the GR action.  See
\cite{DePietri1,Mike1}, for details.

Let us now see how the same can be obtained in 2d.  Consider a
modification of the $BF$ action obtained adding a constraint by means
of a two-form lagrangian multiplier $\lambda$ as follows
\begin{equation}
\label{GRcomeBF}
S_{\rm GR}[B,\omega,\lambda] =-\frac{1}{2}k \int {\rm tr}\,\left[BF -
\lambda\; (BB+{\bf 1}) \,\right] =k \int \left[bf
-\lambda \ (b^2-1) \,\right].
\end{equation}
The equations of motion obtained varying $\omega, B, \lambda$ are,
respectively,
\begin{eqnarray}
dB(x)&=&0,      \label{tre} \\
F(x)&=&2\lambda(x)\tau,     \label{due}\\
-{\textstyle\frac{1}{2}}{\rm tr}[BB]&=&1. \label{uno}
\end{eqnarray}
Eq (\ref{tre}) follows from (\ref{uno}).  Equation (\ref{due}) shows
that the connection $\omega$ is free to take {\em any} value, because
$\lambda(x)$ is arbitrary, precisely as in 2d GR. Equation (\ref{uno})
is the 2d Plebanski constraint.  Inserting it back in the action, and
choosing the value
\begin{equation}
\label{costante}
k=\frac{1}{8\pi G}
\end{equation}
for the coupling constant, we obtain the GR action (\ref{EHeo}).
Recall that in 4d the addition of a constraint frees degrees of
freedom, because the new constraint constrains a constrainer: the
Lagrange multiplier $B$.  Here, something similar happens: the
constraint enlarges the spaces of classical solutions of the the
equation of motion; see \Ref{due}.  However, this does not increase
the number of degrees of freedom because of gauge equivalence.  The
difference is that in 4d the temporal components of the $B$ field are
Lagrange multipliers, while in 2d the $B$ field is simply the
conjugate variable to the spatial component of $\omega$, thus the
Plebanski constraint \Ref{uno} does not constrain a multiplier, and
thus does not add degrees of freedom to the theory.\footnote{We thank
our referees for pointing this out.}

Notice that the Plebanski constraint (\ref{uno}) has two solution:
$B=\pm\tau$, corresponding to the two possible signs of $\det e$.  The
2d theory mimics also the multiplicity of the solutions of the
Plebanski constraint that arises in 4d \cite{LO,DePietri1}.

In 2d the situation is far simpler than in 4d.  The gauge group is
abelian and the theory is topological.  Precisely because of this
simplicity, we are interested in checking whether the Barrett-Crane
technique works in this context.  Thus, we want to quantize 2d GR
following the Barrett-Crane strategy: we write the spinfoam
quantization of the unconstrained $BF$ model, and then impose the
Plebanski constraint (\ref{uno}) as a restriction on the sum over
representations.  Before doing so, however, a crucial remark is
necessary, presented in the following section.

\subsection{Topology of the frame bundle}
\label{topo}

Let us return to the case of a {\em compact\/} 2d space $M$.  Notice
that at first sight there seems to be a contradiction between
(\ref{EH1}) and (\ref{EHeo}): consider the action (\ref{EHeo}) for a
compact manifold $M$.  One could expect the Stokes theorem to
imply that $S_{\rm GR}$ vanishes, since
\begin{equation}
\int_{M} d\omega=\oint_{\partial M} \omega,
\label{sbagliata}
\end{equation}
which vanishes if $M$ is compact and $\partial M=0$.  But $S_{\rm GR}$
is equal to the right hand side of (\ref{EH1}), which in general is
nonvanishing.  How could this be?  The answer is that for a generic
topology of the manifold we cannot define the dyad field $e^i(x)$
globally.  For instance, there is no continuous nonvanishing 1-form
$e^1(x)$ on a sphere.  Thus we must define separate dyad fields
$e_{a}$ on each local chart $a$.  (Here and below $a, b, \ldots$ are
not indices: they are subscripts labeling the chart in which the field
$e_{a}$ is defined.)  These must be related by a $SO(2)$ rotation
$t_{ab}(x)=e^{\psi_{ab}(x)\tau}$ on the overlaps between charts
\begin{equation}
e_{a}= t_{ab}\ e_{b}.
\label{gaugetrprimo}
\end{equation}
In turn, the spin connection one-form is defined only within local
charts, where we call it $\omega_{a}$, and has gauge transformations
\begin{equation}
\omega_{b}=\omega_{a} + d\psi_{ab}.
\label{gaugetr}
\end{equation}
on the overlaps.  Equation (\ref{sbagliata}) is correct only if $e$,
and hence $\omega$, can be defined on the entire $M$.  The only
compact 2d topology where $e^i(x)$ can be defined globally is the
torus, for which (\ref{sbagliata}) is correct; but for the torus
(\ref{sbagliata}) is not in contradiction with (\ref{EH1}), since the
Euler characteristic $\chi$ of the torus is zero.

Consider an arbitrary principal bundle $P(M,SO(2))$ with 2d base
manifold $M$ and structure group $SO(2)$.  Let $\omega$ be a
connection on this bundle and $f$ its curvature.  To compute the left
hand side of (\ref{sbagliata}) using Stokes theorem, we must first
break the integral over $M$ into a sum of integrals over regions $a$,
forming a partition of $M$, and each defined within a single chart.
In each of these regions the connection $\omega$ is represented by a
one-form $\omega_{a}$ and we can apply Stokes theorem.  This gives
\begin{equation}
\label{EHeo2}
\int_{M} f = \sum_{a}\int_{a} f
= \sum_{a}\int_{a} d\omega_{a}
= \sum_{a}\oint_{\partial a} \omega_{a}.
\end{equation}
If we call $e_{ab}$ the boundary separating the regions $a$ and $b$,
with an assigned orientation (say towards the right crossing from $a$
to $b$), this becomes
\begin{equation}
\label{EHeo3}
\int_{M} f = \sum_{ab}\int_{e_{ab}} \omega_a
= \sum_{(ab)} \int_{e_{ab}} (\omega_{a}-\omega_{b})
= \sum_{(ab)} \int_{e_{ab}} d\psi_{ab}\ ,
\end{equation}
where $\psi_{ab}$ is the transition function from the chart in which
the region $a$ is contained to the chart in which the region $b$ is
contained and --here and below-- the notation $(ab)$ indicates
unordered pairs.  Notice that signs are well defined since
$t_{ab}=t_{ba}^{-1}$ implies $\psi_{ab}=-\psi_{ba}$ and therefore
\begin{equation}
\label{EHeo33}
\int_{e_{ab}} d\psi_{ab} = \int_{e_{ba}} d\psi_{ba} .
\end{equation}
Equation (\ref{EHeo3}) shows that the integral of the curvature over
the manifold is independent from the connection, but it depends on the
topology of the bundle, which is coded in the transition functions
$t_{ab}$.  In fact, (\ref{EHeo3}) is ($2\pi$ times) the the Euler
number, a well-known topological invariant that characterizes
the topology of the bundle $P=P(M,SO(2))$,
\begin{equation}
\label{chern}
e(P) = \frac{1}{2\pi}\ \int_{M} f\ .
\end{equation}
The Euler number $e(P)$ is the integral of the first Euler class,\footnote{
In the literature the name of the class usually depends on the structure
group of the bundle; Chern class is used for bundles with structure group $SU(N)$,
and Euler class for bundles with structure group $SO(N)$.
See for instance \cite{Nash}.}
which here is just $f$, over the base manifold. The topology of the
bundle $P$ should not be confused with the topology of the base $M$:
there exist many topologically distinct bundles $P$ for each base
space $M$.  We illustrate this in more detail and give an
explicit example in Section \ref{bundles}.  It can be shown that the
topology of an $SO(2)$ principal bundle over a 2d base space $M$ is
entirely characterized by the Euler number (\ref{chern}).

Let us now return to GR and come to the key point.  In the case of GR,
the gravitational spin connection $\omega$ is a connection on a
principal bundle $P(M,SO(2))$ over spacetime $M$.  However, this is
not an arbitrary bundle: it is (isomorphic to) the frame bundle.  This
is because $\omega$ must satisfy the Cartan equation (\ref{cartan});
$\omega$ is determined by the dyad field $e$, which defines a local
linear map (a ``soldering") between the associated $R^2$ vector bundle
on which $\omega$ acts and the tangent bundle.  This map is smooth and
therefore the bundle on which $\omega$ acts is isomorphic to the
tangent bundle, which in general has nontrivial topology.  Thus,
$\omega$ acts on (a bundle isomorphic to) the tangent bundle, and
therefore it lives in the principal bundle of the local rotations of
the tangent bundle, namely the frame bundle.  We denote the frame
bundle $P_{\rm fr}(M,SO(2))$, or simply $P_{\rm fr}$.  The topology of
the frame bundle is obviously determined by the topology of $M$ and in
general is nontrivial.  Now, in 2d it is well known, and it follows
from (\ref{EH1}), (\ref{EHeo}) and (\ref{chern}), that the frame
bundle is the $P(M,SO(2))$ principal bundle whose Euler number
(\ref{chern}) is equal to the Euler characteristic of the base space
$M$.
That is
\begin{equation}
\label{elegante}
e(P_{\rm fr}) = \chi (M).
\end{equation}
A priori, $BF$ theory can defined over an arbitrary principal bundle;
but if we want to view GR as a modified $BF$ theory, the field
$\omega$ of the $BF$ theory must be a connection on a bundle with the
same topology as $P_{\rm fr}$, namely satisfying
\Ref{elegante}.\footnote{In 2d, the soldering form $e$ has disappeared
from the dynamics, since the action (\ref{EHeo}) depends only on
$\omega$ and a global sign.  However, its ``ghost" remains, since the
vector bundle on which the spin connection acts and the tangent bundle
must have same topology: we can say that they are ``soldered
topologically."} In applying the spinfoam quantization procedure, we
have to be sure that the non-triviality of the bundle is taken into
account, and to choose transition functions $t_{ab}$ satisfying
\Ref{elegante}.  This consideration turns out to play a crucial role
in the quantization.

There is one case which is particularly simple.  This is the case of
genus 1, namely the torus $T_2$.  The tangent bundle of the torus is
trivial: $T(T_2)$ is homeomorphic to $T_2\times R^2$, and the frame
bundle is correspondingly trivial: $P_{\rm fr}(T_2, SO(2))$ is
homeomorphic to $T_2\times SO(2)$.  In fact, we can obviously choose a
global dyad field on the torus, and hence a global spin connection
one-form $\omega$.  Equivalently, the torus admits a flat metric,
hence the curvature can vanish everywhere, and the Euler
characteristic $\chi(T_2)$ is zero.

Accordingly, we separate our analysis of the spinfoam formulation of
2d GR into two steps.  In the following section we discuss the case in
wich the bundle is trivial, where all subtleties related to the global
topology of the frame bundle are absent.  This case is relevant for GR
only when the topology of spacetime is a torus.  In this case, we can
show easily that the Barrett-Crane technique works and yields a finite
and correct partition function.  This result alone, however, is not
very enlightening, since (\ref{Zgenrelreal}) gives simply $Z_{\rm
GR}=2$ for the torus.  Then in Section \ref{nontorus} we discuss the
general case, taking the topology into account, and in Section
\ref{infine} we derive equation \Ref{Zgenrelreal} in the general case.

\section{Trivial principal bundle}\label{torus}

\subsection{Spinfoam quantization of $BF$}\label{Qbf}

Let us recall the standard spinfoam quantization of the $BF$ action
\Ref{BF}.  Formally, the quantum theory is defined by the partition
function
\begin{equation}
\label{ZBF}
Z_{BF}=\int {\cal D}B\,{\cal D}\omega\,e^{\,i\frac{k}{\hbar} \int_M
{\rm tr}
\,[BF]}.
\end{equation}
In order to give meaning to this integral, we define a lattice version
of the theory on a discrete manifold.  Since the theory is
topological, the discretization does not cut off physical degrees of
freedom.  Fix a triangulation $\Delta$ of the manifold.  The
triangulation consists of triangles (2-simplices) meeting at segments
(1-simplices), in turn meeting in points (0-simplices); it is
oriented, with orientation induced by $M$.  Intuitively, we can view
the triangles as flat, with the curvature concentrated on the
0-simplices, as in simplicial Regge geometry \cite{Regge}.  We work
with the cellular complex $\Delta^*$, dual to the triangulation
$\Delta$.  The dual cellular complex $\Delta^*$ is formed by faces $a,
b, c \ldots$ (dual to the 0-simplices).  The two faces $a$ and $b$
meet along an edge that we call $e_{ab}$.  Edges are dual to the
1-simplices.  We orient $e_{ab}$ following the perimeter of $a$
clockwise.  Therefore $e_{ba}$ is the same edge as $e_{ab}$, but with
the opposite orientation.  Edges, in turn, meet at trivalent vertices
$v_{abc}$, which are dual to the triangles of $\Delta$.

\begin{figure}
  \begin{center}
  \includegraphics[height=4cm]{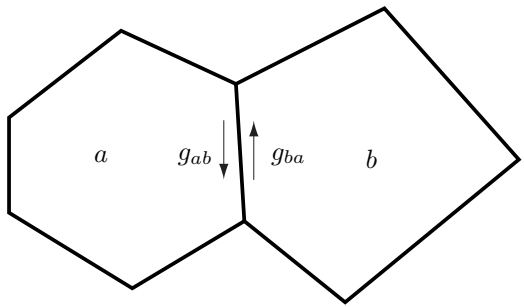}
  \end{center}
  \caption{\label{fig1} Two faces $a$ and $b$ of $\Delta^*$ and the
  group elements $g_{ab}$ and $g_{ba}$ on the edge $e_{ab}$ between
  the two.}
  \end{figure}

We replace the continuous fields $\omega$ and $B$ with discrete fields
on $\Delta^*$, as follows.  The connection $\omega$ is replaced by a
group element $g_{ab}$ associated to each edge $e_{ab}$.  See
Fig.\ref{fig1}.  This is interpreted as the exponential of the one-form field $\frac{k}{\hbar}\omega$, the
``scaled connection", along the edge
\begin{equation}
g_{ab} = e^{\frac{k}{\hbar}\int_{e_{ab}}\omega}.
\label{scaledc}
\end{equation}
Notice that this is not the parallel transport of the metric
connection, which is given by $e^{\int_{e_{ab}}\omega}$.  We shall
comment on the meaning of this choice at the end of Section
\ref{infine}.  We parametrize $SO(2)$ as
\begin{equation}
g_{ab} \equiv e^{\phi_{ab}\tau},
\end{equation}
where $\phi_{ab}\in [0, 2\pi]$.  Since $e_{ba}$ is $e_{ba}$ with the
opposite orientation, we demand that
\begin{equation}
g_{ab}\ g_{ba} = 1 \ ;
\label{eccolo}
\end{equation}
that is
\begin{equation}
\phi_{ab}+\phi_{ba} = 0.
\label{eccolo2}
\end{equation}
The field $B$ is replaced by a variable $B_{a}=b_{a}\tau$ associated
to each face $a$.  The dynamical variables of the lattice theory are
therefore $(g_{ab}, B_{a})$, where $g_{ab}\in SO(2)$, $B_{a}\in so(2)$
and $g_{ab}$ and $g_{ba}$ are related by \Ref{eccolo}.

We now define a group element $g_{a}$ related to the curvature on a
face $a$.  If the face $a$ is bounded by the faces $b_{1}\ldots
b_{m}$,
we define
\begin{equation}
\label{fprodlinks}
g_{a} = e^{\phi_a\tau} \equiv g_{ab_{1}}\ldots g_{ab_{m}} =
e^{(\phi_{ab_{1}}+ \ldots +\phi_{ab_{m}})\tau},
\end{equation}
where $\phi_{a}\in[0,2\pi)$.  Notice that curvature lives on the
faces, which are dual to the 0-simplices of the original triangulation
$\Delta$, which is where we expected curvature to live.  We say that a
discretized connection $g_{ab}$ is flat on $a$ if $g_a={\bf 1}$.  It
is convenient to introduce also the square root of $g_{a}$, defined by
\footnote{From the aesthetic point of view, this is a bit odd,
also because we could take any power $g^a, 0 < a < 1$ of the group
element. We thank our referees for pointing this out.}
\begin{equation}
\label{quadrato}
g^{\scriptscriptstyle{1/2}}_{a}\equiv e^{(\phi_a/2)\tau}.
\end{equation}
{Since we have parametrized the group elements with
$\phi_{a}\in[0,2\pi)$ the last equality defines a square root
uniquely.}

In the limit in which the triangulation is fine and the face $a$ has
area $d^2x$, this satisfies
\begin{equation}
\label{limitecurvatura}
g^{\scriptscriptstyle{1/2}}_{a}={\bf 1}+\frac{1}{2}\, f_{a}\tau\,
d^2x.
\end{equation}

The dynamics of the discrete model is defined by discretizing the
action (\ref{BF}) as
\begin{equation}
S_{BF}[B_{a},g_{ab}]=- \hbar\ \sum_a\ {\rm tr}\, [B_{a}
g^{\scriptscriptstyle{1/2}}_{a}].
\label{BGdiscrete}
\end{equation}
The reason for which it is convenient to use the square root is
technical: it allows us to avoid the doubling of the degrees of
freedom introduced by the discretization, as we shall see in a
moment. {An alternative technique for dealing with this problem
was developed by Freidel and Louapre in \cite{Freidel3}.}

This action approximates the continuous action in the limit of fine
triangulations and reproduces the symmetries of the continuous action
(see \cite{Freidel2} for the 3d case.)  The action is invariant under
local gauge transformations generated by a gauge parameter
$\lambda_{abc}$, which sits on vertices $v_{abc}$.  The discretized
connection transforms as
\begin{equation}
g_{ab}\rightarrow \lambda_{abc} \ g_{ab}\  \lambda^{-1}_{abd} \ ,
\end{equation}
where $v_{abc}$ and $v_{abd}$ are the vertices where the edge $e_{ab}$
begins and ends.

We can now give a precise definition to the formal functional integral
(\ref{ZBF}) as
\begin{equation}
\label{ZBcontinuo}
Z_{BF} =  \int \prod_{(ab)}\ dg_{(ab)}\;\prod_a\  dB_{a}
\ e^{\frac{i}{\hbar}S[B_{a},g_{ab}]}  =  \int  \prod_{(ab)}dg_{(ab)}\;
 \prod_{a}dB_{a}
\ e^{-i \sum_{a} {\rm tr}\,[B_{a} g^{1/2}_{a}]} .
\end{equation}
Here
\begin{equation}
dg_{(ab)} = dg_{ab}\ dg_{ba}\ \delta(g_{ab}g_{ba})
\label{delta}
\end{equation}
and $dg_{ab}$ is the Haar measure of $SO(2)$.  This is just the
normalized measure of an angle, $dg_{ab}\equiv d\phi_{ab}/2\pi$.    We
take for $dB_{a}$ the Lebesgue measure on the $so(2)$ algebra, that is
$dB_{a}\equiv db_{a}$.  This defines quantum $BF$ theory.  Let us now
compute the partition function explicitly.

The integrals over $dB_{a}$ can be easily performed.
Using
\begin{equation}
- {\rm tr} \,[Bg]=- {\rm tr} \,\left[b
\left(\begin{array}{cc} 0 & 1 \\ -1 & 0\end{array}\right)
\left(\begin{array}{cc} \cos \phi & \sin \phi \\ -\sin
\phi &
\cos \phi \end{array}\right)\right]
= 2b \sin\phi  \ ,
\end{equation}
we have
\begin{equation}
\label{ZBF3}
\int dB_{a}\; e^{-i\ {\rm
tr}\,[B_{a}\,g^{\scriptscriptstyle{1/2}}_{a}]}=
\int db_{a}\; e^{2ib_{a} \sin(\phi_{a}/2)}=
2\pi\, \delta(2\sin(\phi_{a}/2))=2\pi\, \delta(\phi_{a}) =
\delta(g_{a})\ ,
\end{equation}
where $\delta(g)$ is the delta function on the group, defined with the
Haar measure.  (Had we used $g_{a}$ instead of
$g_{a}^{\scriptscriptstyle{1/2}}$ in the action, we would have
obtained $
2\pi\delta(\sin\phi_{a})=2\pi\left(\delta(\phi_{a})+\delta(\phi_{a}-\pi)\right)
= \delta(g_{a})+\delta(-g)$.  See \cite{Freidel3} for an alternative
solution to the same problem.)  Using this, we
obtain,
\begin{equation}
\label{ZBcontinuobis}
Z_{BF}=  \int  \prod_{(ab)}\ dg_{(ab)}  \prod_{a} \ \delta(g_{a}),
\end{equation}
which is the standard expression for the partition function of the
$BF$ models.

The spinfoam sum is obtained using the Plancherel expansion the
$\delta$ function on the group in irreducible representations.  In our
simple abelian case the $SO(2)$ irreducible representations are
labelled by an integer $n$; they are one-dimensional and
their character is $\chi^{(n)}(g)= e^{in\phi}$.  In other words, the
Plancherel decomposition is here the standard Fourier expansion of the
delta function
\begin{equation}
\delta(\phi)=\frac{1}{2\pi} \sum_n e^{in\phi},
\end{equation}
Using this, we can write
\eqa
\label{ZBF4}
Z_{BF}&=&\int \prod_{(ab)}\ \frac{d\phi_{(ab)}}{2\pi}\ \ \prod_a \
\sum_n \,e^{in\phi_{a}}.  \neqa We can exchange the sum with the
integrals by writing this expression as a sum over all the possible
assignments $\{n_{a}\}$ of a representation $n_{a}$ to each face $a$
\begin{equation}
\label{ZBF5} Z=\sum_{\{n_a\}}\int
 \prod_{(ab)}\ \frac{d\phi_{(ab)}}{2\pi}\ \ e^{i\sum_a n_{a}\phi_{a}}.
\end{equation}
Each edge bounds exactly two faces.  Using \Ref{fprodlinks}, this
becomes
\begin{equation}
\label{ZBF54}
Z=\sum_{\{n_a\}}\int  \prod_{(ab)}\
\frac{d\phi_{(ab)}}{2\pi}\ e^{i\sum_{ab} n_{a}\phi_{ab}}.
\end{equation}
Using \Ref{eccolo2}, this becomes
\begin{equation}
\label{ZBF55}
Z=\sum_{\{n_a\}}\int  \prod_{(ab)}\
\frac{d\phi_{(ab)}}{2\pi}\  e^{i\sum_{(ab)} (n_a-n_b)\phi_{ab}}\ ,
\end{equation}
where the sum is over unordered pairs.  All integrals decouple.
They can be performed using the abelian equivalent of the formula for
the orthogonality of characters
\begin{equation}
\label{ortho}
\int_0^{2\pi}\frac{d\phi}{2\pi}\,e^{-in_{a}\phi}\,
e^{in_{b}\phi}=\delta_{n_{a}n_{b}}.
\end{equation}
The multiple sum reduces to a single sum of the identity over one
index only,
\begin{equation}
\label{tutto11}
Z_{BF}=\sum_n \,1.
\end{equation}

Thus the partition function diverges badly, as does the partition
function of $BF$ theory in 4d.  Intuitively, this divergence is due to
the infinite volume of the original algebra-valued variable $B(x)$, or
equivalently by the presence of a translational gauge symmetry $B\rightarrow B + c$
for any constant field $c$, and
could formally be eliminated by dividing $Z$ by the infinite volume of
the algebra \cite{Freidel2}.  As we shall see, this
divergence disappears in the GR case, and therefore we do not worry
about it here.  On the other hand, the partition function is formally
triangulation independent, as we expect from a topological quantum
field theory with no local degrees of freedom.

The sum \Ref{tutto11} is an extremely simplified spinfoam model: it is
defined on a 2-complex, where faces are colored by representations of
the gauge group.  The ``abelian Clebsch-Gordon conditions" force
representations to match across each edge, hence they force all
representations to be equal.  The partition function of the model is
then given by the weighted sum over all allowed colorings.  The weight
of each coloring is 1 for all representations.

\subsection{Quantum GR on the torus}

Let us now come to GR. As mentioned in the introduction, our strategy
is to view GR as a $BF$ theory plus the additional Plebanski
constraint (\ref{uno}), and to search for its quantum theory by
enforcing this constraint into the spinfoam sum that defines quantum
$BF$ theory.  In order to do so, we have to identify the quantity
representing the field $B$ in the spinfoam sum.

As in 4d, it is the representation $n_{a}$ associated to the face $a$
that represents the field $B_{a}$ on the face.  Indeed, observe that
in equation (\ref{ZBF3}) we have integrated over $B_{a}$ to get the
delta function; and then in equation (\ref{ZBF4}) we have expanded
back the delta function as a sum over $n_{a}$.  That is, we have used
the fact that
\begin{equation}
\int db_{a}\; e^{i\,b_{a}\,\phi_{a}}= 2\pi\,\delta(\phi_{a}) \ =
\sum_{n_{a}}\ e^{in_{a}\phi_{a}}.
\end{equation}
The equality is correct because the delta is defined on a compact
domain: $\phi_{a}\in[0,2\pi]$.  (In fact, this equality expresses the
reason for which a physical quantity conjugate to a variable living in
a compact domain is quantized.)  The discrete variable $n_{a}$ is
therefore the quantized version of the continuous variable $B_{a}$.
Hence, we translate the classical constraint on $B_{a}$ into a
constraint on $n_{a}$.  That is, we implement the Plebanski constraint
\Ref{uno} as a restriction on the representations summed over, as in
done in the Barrett-Crane models.  We can mimic exactly the 4d
procedure by identifying the continuous field $B\in so(2)$ with the
$so(2)$ generator $J$, as we identify angular momentum with the
generator of the rotation group.  Consider the quadratic Casimir $C=
-{\frac{1}{2}}{\rm tr} [JJ]$.  Under the identification $B
\leftrightarrow J$ the Plebanski constraint \Ref{uno} reads
\begin{equation}
C=-{\frac{1}{2}}{\rm tr} [JJ]=-{\frac{1}{2}}{\rm tr} [BB]=1.
\end{equation}
In the representation $n$, the generator $J$ is the matrix $n\tau$ (we
are still working with representations of the algebra over the real
numbers, with the generator given by the expression \Ref{tau}) and
the Casimir $C$ takes the value
\begin{equation}
C=-{\frac{1}{2}}{\rm tr} [n\tau n \tau ]=n^2.
\end{equation}
Hence the Plebanski constraint becomes
\begin{equation}
n^2=1.
\label{Pln}
\end{equation}
We enforce the Plebanski constraint (\ref{uno}) by restricting the
spinfoam sum \Ref{ZBF54} to the representations satisfying \Ref{Pln}.
Then the infinite sum reduces to two terms, giving
\begin{equation}
\label{tutto1!}
Z_{\rm GR}[T_2]=\sum_n \delta_{n^2,1} = 2.
\end{equation}
where we have indicated explicitly that the theory is defined on a
torus.

First, notice that the model becomes finite. {At the end of
the previous section we mentioned the fact that the symmetry under
translations of the $B$ field is responsible for the divergence of
the partition function.  The Plebanski constraint \Ref{uno} fixes
this gauge, yielding a finite partition function.}

{Notice that also in 4d the Plebanski constraint fixes part of the
gauge symmetry, since the gauge symmetry group of $BF$ theory is
larger than the one of GR. The interpretation of the the
divergences of 4d spinfoam gravity theory, however, is
controversial.  Some models are most probaly finite
\cite{PR,finiteness,finiteness2} (see \cite{Perez2,book}); for
these, the analogy with the 2d case presented here is complete.
But it has also been argued that the remaining divergences, that
appear in other 4d spinfoam gravity models, are to be expected,
because they reflect the rest of the gauge invariance (see
\cite{Freidel2} for the relevant analysis in the 3d case).  If so,
an extra gauge fixing has to be performed to obtain finite
results. We do not enter in this debate here.  In any case, the
simple model discussed here illustrates how full gauge fixing is
the condition for finiteness.}

Second, at first sight
the result $Z_{\rm GR}[T_2]=2$ may seem trivial, but notice that it
agrees precisely with (\ref{Zgenrelreal}), for the following reason:
we have assumed the bundle to be trivial, but in GR the bundle must be
the frame bundle.  The only case in which the frame bundle is trivial
is the torus, for which $e(P_{\rm fr})=\chi(T_2)=0$.  Thus, the result
we obtain is only valid for the torus, and therefore it agrees with
(\ref{Zgenrelreal}).  The result is therefore finite and correct.

However, to be convinced that this works in the general case
we have to deal with arbitrary bundle topologies.  For this, we need
to define topologically nontrivial simplicial bundles; this is what we
do in the following section.

\section{Nontrivial bundle topology}\label{nontorus}\label{math}

Our aim is now to give a useful definition of a discrete 2d bundle,
that is, a bundle over a cellular complex, defined with discrete
topological data.  This can be done in several ways; here we discuss a
definition that allows a simple discretization of a field theory on a
nontrivial continuous bundle.  To motivate our definitions, we proceed
by constructing the discrete bundle starting from an actual
discretization of the continuous one.  In the Appendix we present a
different and more detailed construction, which should turn out to
better match with the GFT formulation of the theory and to be more
appropriate to the extension to higher dimensions.

\subsection{Bundles}
\label{bundles}

Let us begin by recalling some simple notions of differential
geometry.  A principal bundle $P(M,G)$, where $G$ is a Lie group and
$M$ is a compact manifold, can be uniquely reconstructed (as an
equivalence class of coordinate bundles) by covering $M$ with
topologically trivial open subsets $U_a$ and giving a set of
transition functions
\begin{equation}
t_{ab}: U_{a}\cap  U_{b} \to G.
\end{equation}
The transition functions must to satisfy three conditions:
\begin{eqnarray}
t_{\alpha\alpha}(x)&=&{\bf 1}, \no
t_{\alpha\beta}(x)\ t_{\beta\alpha}(x)&=&{\bf 1}, \no
t_{\alpha\beta}(x)\ t_{\beta\gamma}(x)\ t_{\gamma\alpha}(x)&=&{\bf 1},
\label{coci}
\end{eqnarray}
for all points $x\in M$ where the functions are all defined.  The last
condition is the {\it cocycle} condition. A gauge transformation
is defined to be a collection of maps
\begin{equation}
\lambda_{a}: U_{a} \to G.
\end{equation}
acting on the transition functions in the following way,
\begin{equation}
\label{gaugetrans0}
{t_{\alpha\beta}}(x) \longmapsto\lambda^{-1}_\alpha(x)\
{t_{\alpha\beta}}(x)\ \lambda_\beta(x).
\end{equation}
Gauge-transformed transition functions define the same bundle.  A
bundle is characterized by a gauge equivalence classes of transition
functions satisfying \Ref{coci}.

A connection $\omega$, defined globally, is represented by a
connection one-form $\omega_{a}$ on each $U_a$, with values in the Lie
algebra of $G$, where, on each overlap $U_{a}\cap U_{b}$,
\begin{equation}
\label{gaugetrasfaperti}
\omega_\alpha=\text{Ad}_{t^{-1}_{\alpha\beta}}
\,
\omega_\beta + t^{-1}_{\alpha\beta}\,d t_{\alpha\beta}\ ,
\end{equation}
where $\text{Ad}$ is the adjoint representation.  The curvature two
form is $f_{a}=d\omega_{a}+\omega_{a}\wedge\omega_{a}$.  The
information about the topology of $P(M,G)$, which is coded in the
transition functions, can be extracted from integrals of polynomials
in the curvature, called characteristic classes, of an arbitrary
connection, which represent topological invariants.

In the case where $G=SO(2)$, \Ref{gaugetrasfaperti} reduces to
\Ref{gaugetr}, the curvature is gauge invariant and therefore is a
globally defined two-form $F=f\tau$.  If $M$ is a 2d connected compact
and orientable manifold, the only interesting characteristic class is
the first Euler class, whose integral gives the Euler number
\Ref{chern}.  Since $df=0$,
\begin{equation}
\label{class}
e(P) = \frac{1}{2\pi}\int_M f
\end{equation}
is an element of the second de Rham cohomology group,
$H^2(M,\mathbb{Z})\sim \mathbb{Z}$.  The inequivalent principal
bundles $P(M,SO(2))$ are therefore classified by the integer $e(P)$.
As discussed above, the frame bundle of $M$ is a $P(M,SO(2))$ bundle
with $e(P)=\chi(M)$.

Let us consider the sphere $M=S_2$ as an example.  Let
$\theta\in[0,\pi],\ \phi\in[0,2\pi]$ be the standard polar coordinates
of the sphere.  We can choose two open sets, the (slightly extended)
northern and the southern hemisphere, $U_N$, defined by $\theta\in[0,
\pi/2+\epsilon]$, and $U_S$ defined by $ \theta\in
[\pi/2-\epsilon,\pi]$.  The overlap region is the 2$\epsilon $-wide
ribbon around the equator.  It is not difficult to see that any
transition function is gauge equivalent to a transition function of
the form $t_{ab}(\theta,\phi)=e^{q\phi\tau}$, where $q$ is an integer.
Notice that $q$ counts the number of times the equator wraps around
the gauge group.  A connection\footnote{The connection (\ref{wuyang})
is the one encountered in the Dirac monopole: $q$ is the magnetic charge times the
electric charge, and the condition that $q$ is integer is Dirac's
quantization of the magnetic charge.} can be written in the form
\cite{Naka1}
\begin{equation}
\label{wuyang}
\omega_N=\frac{q}{2}(1-\cos\theta)\ \ d\phi, \hspace{0.5cm}
\omega_S=-\frac{q}{2}(1+\cos\theta)\ \  d\phi.
\end{equation}
On the equator, the two one-forms are related by
\Ref{gaugetrasfaperti}, which takes the form $\omega_N=\omega_S+ q
d\phi$.  The Euler number \Ref{class} is
\begin{equation}
 e(P) = \frac{1}{2\pi}\int_{S^2} f = \frac{1}{2\pi}\int_{\text{\rm
 equator}}[\omega_N-\omega_S]= \frac{1}{2\pi}\int_0^{2\pi}q\, d\phi=q.
\end{equation}
For $q=0$ we have the trivial bundle $S_2\times SO(2)$, which admits
the flat connection (\ref{wuyang}).  For $q=1$ we have the principal
bundle associated to the Hopf vector bundle (the complex line bundle
over the projective plane).  Since the Euler number of the sphere is
$\chi(S^2)=2$, for $q=2$ we have the frame bundle.  To see this,
notice that the commonly used dyad $e^1=d\theta$, $e^2=\sin\theta
\,d\phi$ is ill-defined at the two poles.  We can get a dyad which is
well
defined at the poles by rotating it in the two hemispheres
\begin{eqnarray}
\label{dyade}
e^1_N\ =\ \cos\phi\ d\theta\, -\, \sin\phi\ \sin \theta\ d\phi,
&\hspace{0.5cm}&
e^2_N\ =\ \sin\phi\ d\theta\, +\, \sin\phi\ \sin \theta\ d\phi,
\\
e^1_S\ =\ \cos\phi\ d\theta\, +\, \sin\phi\ \sin \theta\ d\phi,
&\hspace{0.5cm}&
e^2_S\ =\ -\sin\phi\ d\theta\, +\, \sin\phi\ \sin \theta\ d\phi.
\end{eqnarray}
The corresponding connection, that satisfies the Cartan equation
\Ref{cartan}, is (\ref{wuyang}) with $q=2$.  It is easy to get
convinced that if $q$ is different from 2, then the connection
(\ref{wuyang}) does not admit a dyad satisfying \Ref{cartan}.

\subsection{Discrete bundles}\label{db}

Let us now define a bundle over a 2d discretized manifold.  Consider a
2d manifold $M$ and a cellular decomposition $\Delta^*$ of $M$.
$\Delta^*$ is formed by 2d faces $a,b,c,\ldots$, which partition $M$.
The faces meet along 1-dimensional edges $e_{ab}$, which, in turn,
meet in vertices $v_{abc}$.  (For simplicity we assume here that
vertices are trivalent.)  Associate to each face $a$ an open subset
$U_{a}$ of $M$, containing $a$.  Let a bundle be defined by transition
functions $t_{ab}$ on the overlaps.  Notice that the edge $e_{ab}$ is
contained in the overlap $U_{a}\cup U_{b}$.  The topological
information on the bundle is entirely contained in the restrictions
of the transition functions to the edges.  Let $s\in[0,1]$ be a
coordinate on the edge $e_{ab}$ and let us call $t_{ab}(s)$ this
restriction.  That is, we consider functions
\begin{equation}
t_{ab}: e_{ab} \to G.
\end{equation}
These transition functions satisfy three conditions:
\begin{eqnarray}
t_{\alpha\alpha}(s)&=&{\bf 1},  \label{singlo}\\
t_{\alpha\beta}(s)\ t_{\beta\alpha}(s)&=&{\bf 1}, \label{duppo}\\
t_{\alpha\beta}(s)\ t_{\beta\gamma}(s)\ t_{\gamma\alpha}(s)&=&{\bf 1},
\label{coci2}
\end{eqnarray}
for all points where the functions are all defined.  Notice that now
the cocycle condition holds at isolated points: the vertices
$s=v_{abc}$.  A gauge transformation is a collection of continuous
maps from the perimeter of each face $a$ to $G$, such that the image
of $\lambda_{a}$ is contractible (because it must be possible to
extend it continuously on the entire face.)  It acts on the transition
functions in the following way,
\begin{equation}
\label{gaugetrans}
t_{\alpha\beta}(s) \longmapsto \lambda^{-1}_\alpha(s)\
t_{\alpha\beta}(s)\ \lambda_\beta(s).
\end{equation}
A principal bundle $P(\Delta^*, G)$ is defined by the transition
functions (\ref{coci2}) up to the gauge transformations
\Ref{gaugetrans}.\footnote{The cohomological structure $d^2=0$
determined by $d:\lambda_\alpha \longmapsto t_{\alpha\beta}=
\lambda^{-1}_\alpha \lambda_\beta $ and $ d: t_{\alpha\beta}
\longmapsto \rho_{abc}= t_{ab}\, t_{bc}\, t_{ca}$ is the same as in
the continuous case, but restricted to $\Delta^*$.} As suggested by
\Ref{EHeo3}, the key quantity of interest is the {\em variation} of
the transition function along each edge.  Let us therefore define the
quantity
\begin{equation}
\label{gaugetransnonab}
n_{ab}= \frac{1}{2\pi} \int_{e_{ab}}ds\ \ t^{-1}_{\alpha\beta}(s)\
\frac{d}{ds} t_{\alpha\beta}(s)
\end{equation}
associated to each edge.  Restricting to the case $G=SO(2)$ we are
interested in, this can be expressed as
\begin{equation}
\label{gaugetrans22}
n_{ab}= \frac{1}{2\pi}
\int_{e_{ab}} d\psi_{ab}(s),
\end{equation}
where $t_{ab}(s)=e^{\psi_{ab}(s)\tau}$.  Notice that $n_{ab}=n_{ba}$,
because reversing the order of the cells changes the orientation of
the edge but also the sign of $\psi_{ab}(s)$, because of
(\ref{duppo}).  A gauge transformation in the open set $a$ adds a real
number $\lambda_{ab}$ to each $n_{ab}$.  This number is the variation
of the gauge transformation along $e_{ab}$, and can be written as
$\lambda_{ab}=\lambda^i_{a}-\lambda^j_{a}$, where $i$ and $j$ label
the two vertices bounding $e_{ab}$.  The gauge transformation can be
written as
\begin{equation}
\label{gaugetransd}
n_{\alpha\beta} \mapsto n_{\alpha\beta} + \lambda^i_{a} -
\lambda^j_{a} + \lambda^j_{b} - \lambda^i_{b}.
\end{equation}

This structure can be simplified using a partial gauge fixing.
Because of the cocycle condition (\ref{coci2}) it is always possible
to choose a gauge transformation \Ref{gaugetrans} such that the
transformed $t_{\alpha\beta}(s)$ is the identity at the vertices,
namely $t_{\alpha\beta}(0)=t_{\alpha\beta}(1)=1$.  Denote this gauge
the ``edge gauge".  Then the cocycle condition is trivially satisfied.
In this gauge the transition functions define maps $t_{\alpha\beta}:
S_{1}\to G$.  A gauge transformation deforms these maps smoothly:
therefore what is relevant is the homotopy class of these maps.  The
bundle can be nontrivial only if $\pi_{1}(G)$, the first homotopy group
of $G$, is nontrivial.  Notice that for $G=SO(2)$ in this gauge
\Ref{gaugetrans22} is precisely the number of times the transition
function of the edge wraps around the group, and is an integer. It
is an element of $\pi_{1}(SO(2))$.
Within the edge gauge, we can still gauge transform the integers
$n_{ab}$ by means of a gauge transformation that is the identity on
the vertices, but the quantities $\lambda^i_{a}$ in \Ref{gaugetransd}
must be integers (more precisely, elements of $\pi_{1}(SO(2))$.)

The bundle is thus uniquely characterized by the discrete set of
integers $n_{ab}$ (more precisely, elements of $\pi_{1}(SO(2))$) and
the gauge transformations (\ref{gaugetransd}).  It is clear that the
only invariant quantity is
\begin{equation}
\label{kdis}
e(P)=\frac{1}{2}\sum_{ab} n_{\alpha\beta},
\end{equation}
which can be recognized as the Euler number of the bundle.

We take the above construction as a motivation, and {\em define} the
``discrete bundle" $P(\Delta^*,SO(2))$, namely an $SO(2)$ bundle over
an abstract 2d cellular complex $\Delta^*$, by the assignement of an
element of $\pi_{1}(SO(2))$ to each edges $e_{ab}$ of $\Delta^*$, up
to the gauge transformations \Ref{gaugetransd}.  The extension of this
definition to different groups is straightforward.  The extension to
higher dimensional manifolds will be considered elsewhere.

\section{Quantum $BF$ and GR on an arbitrary
topology}\label{infine}

On a nontrivial bundle, we associate a distinct connection one-form
$\omega_{a}$ to each open set $U_{a}$.  On a cellular manifold, we
associate a distinct connection one-form $\omega_{a}$ to each face $a$
(we have assumed that each face is included in an open set.)  In
discretizing the field theory, we previously considered a single group
element \Ref{scaledc} associated to each edge $e_{ab}$, separating the
face $a$ from the face $b$; this is obtained exponentiating the
(scaled) connection.  If the bundle is nontrivial, we must consider,
instead, {\em two} of these group elements, one defined by the
connection $\omega_{a}$ and the other defined by the connection
$\omega_{b}$.  Let us denote them $g_{ab}$ and $g_{ba}$:
\begin{equation}
g_{ab} = e^{\frac{k}{\hbar}\int_{e_{ab}}\omega_{a}}, \hspace{2em}
g_{ba} = e^{\frac{k}{\hbar}\int_{e_{ba}}\omega_{b}}.
\label{scaledcab}
\end{equation}
These two group elements are related by the transition functions.
From \Ref{gaugetr} and \Ref{gaugetrans22} we have
\begin{equation}
g_{ab}\ g_{ba} = e^{\frac{k}{\hbar}2\pi n_{ab}\tau},
\label{gabgba}
\end{equation}
which replaces \Ref{eccolo} in the general case.

We can now proceed with the quantization precisely as we did for the
torus.  The only difference is that the relation between $g_{ab}$ and
$g_{ba}$ is given by \Ref{gabgba} instead than by \Ref{eccolo}.
Consequently, the definition of the measure \Ref{delta} must be
replaced by
\begin{equation}
dg_{(ab)} = dg_{ab}dg_{ba}\ \delta(g_{ab}g_{ba}
e^{-\frac{k}{\hbar}2\pi n_{ab}\tau}),
\label{deltanuova}
\end{equation}
Repeating the quantization precisely as in Section \ref{torus},
we get again to equation \Ref{ZBF54}, but now the relation
between $\phi_{ab}$ and $\phi_{ba}$ is determined by \Ref{gabgba},
namely is
\begin{equation}
\phi_{ab}+ \phi_{ba} = \frac{k}{\hbar}\ 2\pi\ n_{ab}
\label{phiabgphiba}
\end{equation}
instead of (\ref{eccolo2}).  Therefore \Ref{ZBF55} is replaced by
\begin{equation}
\label{ZBF55rep}
Z=\sum_{\{n_a\}}\int \prod_{(ab)}\
\frac{d\phi_{(ab)}}{2\pi}\ e^{i\sum_{(ab)}\left(
(n_a-n_b)\phi_{ab}+n_{b}\frac{k}{\hbar} 2\pi n_{ab})\right)}.
\end{equation}
performing the integration, and using \Ref{kdis}, this gives
\begin{equation}
\label{nuovo}
Z =\sum_n  e^{in\frac{k}{\hbar}2\pi\sum_{(ab)} n_{ab}} =
\sum_n  e^{in\frac{2\pi k}{\hbar} e(P)}.
\end{equation}

In the case of $BF$ theory, we can perform the sum, giving
\begin{equation}
\label{nuovo2}
Z_{BF} = \delta(k/\hbar \ e(P)).
\end{equation}
That is, the partition function of $BF$ theory vanishes unless
$e(P)=0$, namely unless the bundle is trivial.  In a sense, this mean
that if the bundle is nontrivial $BF$ theory does not exist.  The
result can be related to what happens in the classical theory, because
on a nontrivial bundle the $BF$ equation \Ref{em2} has no solution,
since there is no flat connection on a nontrivial bundle.  If the
bundle is trivial, we recover the divergence mentioned above.

In the case of GR, recall that we must: (i) choose the value
\Ref{costante} for the coupling constant $k$; (ii) fix the bundle to
be the frame bundle, namely impose \Ref{elegante}; and (iii)
impose the Plebanski constraint \Ref{Pln}. Using this, we
have immediately
\begin{equation}
\label{etvoila}
Z_{\rm GR}[M]\ =\ \sum_n \ \delta_{n^2,1} \ e^{in\frac{1}{4\hbar
G}\chi(M)}\ =\
2\ \cos\left(\frac{\chi(M)}{4\hbar G}\right).
\end{equation}
This is precisely the result for the partition function of 2d
riemannian quantum GR, equation \Ref{Zgenrelreal}, as desired.  This
result has been obtained with a finite calculation, where no
divergences or ill-defined quantities appear.

\vskip.5cm

We close this section with a few comments.

First, the same result can be obtained also by directly discretizing
and quantizing the action \Ref{GRcomeBF}.  The discretization of
\Ref{GRcomeBF} can be obtained by adding one term to the discrete
action (\ref{BGdiscrete})
\begin{equation}
S_{BF}[B_{a},g_{ab}]=- \hbar\ \sum_a\ {\rm tr}\, [B_{a}
g^{\scriptscriptstyle{1/2}}_{a}-\lambda_{a}(B_{a}B_{a}+{\rm 1})],
\label{BGdiscretec}
\end{equation}
and all the integrals in the definition of the partition function can
be easily performed.
Here we have preferred to enforce the constraint ``by hand" as a
constraint on the representations in order to mimic the 4d procedure.

Second, to obtain the complex partition function \Ref{PartGR},
corresponding to a fixed orientation of the spacetime manifold, it
is sufficient to restrict the determinant of the dyad to be positive.
This restriction of the global orientation is not straightforward in
higher dimensions (see \cite{LO}), but in this 2d case can be easily obtained by replacing the action
\Ref{GRcomeBF} with
\begin{equation}
\label{GRcomeBF2}
S_{\rm GR}[B,\omega,\lambda] =-\frac{1}{2}k \int {\rm tr}\,\left[BF -
\lambda\; (B\tau+{\bf 1}) \,\right] =k \int \left[bf
-\lambda \ (b-1) \,\right].
\end{equation}
In this way, the only solution of the Plebanski
constraint is $B=+\tau$, i.e. $b=+1$, equivalent to $n=+1$ at the
quantum level.

Finally, the reader may wonder what is the meaning or the rationale
for the factor $k/\hbar$ that multiplies the gravitational connection
in the definition \Ref{scaledc} of the group elements used as
variables for the discretized theory.  At first sight, one may think
that this factor can be simply absorbed in the definition of the units
in which the connection is measured.  Notice that $k/\hbar=1/8\pi
G\hbar$ is dimensionless in 2d.  We can certainly absorbe this factor
in the field by re-scaling $\omega$ and defining
$\tilde\omega=(k/\hbar) \, \omega$.  However, if we multiply a
connection by a real number, we do not obtain a connection.  More
precisely, if we insist that $\tilde\omega$ be a connection, we have
to modify the gauge trasformations, and thefore, ultimately, change
the ``size" of the gauge group, which shows up in \Ref{gaugetrans22}.
In other words, the quantum theory does depend on a dimensionless
number which expresses precisely the ratio between the dimension of
the phase defined by the action and the dimension of the gauge group.

\section{Summary and perspectives}\label{conclusion}

Riemannian general relativity in 2d can be written as a $BF$ theory
plus a Plebanski-type constraint.  This is analogous to what happens
in 4d.  The quantization can be performed imposing the constraint in
the spinfoam theory as a restriction of the representations summed
over.  The partition function of $BF$ theory diverges on a trivial
bundle and vanishes on nontrivial bundles.  The partition function of
GR, on the other hand, is well-defined for all topologies of the
spacetime manifold, there are no divergences, and agrees with the one
obtained from formal manipulations of the functional integral.  This
result can be taken as indirect support to the viability of similar
techniques used in 4d quantum gravity.

Care should be taken concerning the bundle topology.  The topology of
the bundle on which the gravitational connection is defined is the
frame bundle.  It is determined by the topology of spacetime and in
general it is nontrivial.  In the discretization of the continuous
theory, we need to take the topology of the bundle into account.  This
topology can be coded by integers (elements of $\pi_{1}(SO(2))$)
associated to the edges of the cellular complex used in the
discretization.  These data define a nontrivial $SO(2)$
principal bundle over a simplicial complex.  The spinfoam theory must
be defined on this nontrivial discrete bundle.

We close with some general considerations on possible developments.
{Bundle topology, and higher dimensional homotopy groups, might play a
role also in higher dimensions.  It may be possible to refine the
higher dimensional spinfoam models introducing the additional data
encoding the fact that we are dealing with a connection on the
frame bundle, which is topologically non-trivial for non-trivial
spacetime manifolds.  In the 3d case we expect the $\pi_2(G)$
homotopy group to provide the relevant data.  Notice that
nontriviality appears only if we use $SO(3)$, since any $SU(2)$
bundle in 3d is trivial.}

{In 4d, it is reasonable to expect that the relevant homotopy
group is $\pi_3(SO(4))$.  In the case, the topological data entering
the action are likely to be insufficient to characterize the frame
bundle.  But a detailed analysis is needed before drawing a
conclusion.  In 4d, GR has local degrees of freedom, and global issues
may be of lesser importance.  However, when the definition of the
theory includes a sum over topologies, topological dependent factors
might sensibly affect the sum.}  Notice for example that spacetime
topology is not fixed in the group field theory (GFT) approach to the
spinfoam formalism; see \cite{Daniele1,Perez2}.

If quantum gravity is formally defined as a path integral for the
gravitational connection field $\omega$, one might also consider the
possibility of treating the topology of the bundle as a variable in
the quantum theory, and of summing over bundle topologies in a
sum-over-histories formulation.  Here, we have considered the bundle
topology fixed to the value imposed by the Cartan equation
(\ref{cartan}).  In the GFT formalism described in the appendix, on
the other hand, it appears that the quantities encoding the bundle
topology can only be added as dynamical variables to be summed over.

Finally, we have not discussed transition amplitudes.  In a
topological quantum field theory, these can be obtained by computing
the partition functions with fixed boundary conditions \cite{Atiyah1}.
The Euler characteristic of a 2d compact manifold with $b$ boundaries
is given by $\chi(M)=2-2g-b$.  We expect that the setting discussed
here could be extended to the case with boundaries, with boundary data
in the $n=1$ representation of $SO(2)$, and transition amplitudes
weighted by $e^{\frac{i}{4 \hbar G} \chi(M)}$.  For a manifold with
genus $g$ and $b$ boundaries, the only nontrivial transition function
should have the structure
\begin{equation}
W_{\rm GR}\sim e^{\frac{4\pi i}{\hbar G}(2-2g-b)}.
\end{equation}
We will discuss these amplitudes, together with the GFT that
generates them, elsewhere.

Among other issues that have not yet been studied in this context, we
signal the lorentzian theory, matter and particle couplings (see
\cite{Freidel3,particles}).

In spite of its simplicity 2d gravity proves to be a model where there
is still much to learn.

\vspace{1cm} \noindent {\bf Acknowledgments} \\ Thanks to Dario
Benedetti, Matyas Karadi and Etera Livine for useful discussions.
Thanks to Andrea Sambusetti for many clarifications on principal
bundles and the Gauss-Bonnet theorem.  SS thanks Rome's Physics
Department and INFN's group TS11 for support.

\section*{Appendix}

Here we describe a second version of the construction of a discrete
bundle.  In Section \ref{db} we assumed that each {\em face} of the
cellular complex $\Delta^*$ was contained in a single chart.  Here we
assume that each {\em triangle} of the triangulation $\Delta$ is
contained in a single chart.  This construction is equivalent but
slightly more complicated than the one given in \ref{db}; it has the
advantage that it matches very naturally the GFT formalism.
Furthermore, it should be more adapted to higher dimensions.  Only in
2d the faces of $\Delta^*$ have the same dimension as spacetime.  In
$n$ dimensions, we can relate the $n$-simplices of the simplicial
decomposition $\Delta$ of the manifold with the ($n$-dimensional) open
sets where the local charts are defined.

Assume that each triangle $i,j, k, \ldots$ of a triangulation $\Delta$
lies within a single chart.  Denote $s_{ij}$ the oriented segment
separating the triangles $i$ and $j$.  Consider the restriction of the
transition functions to these segments.  This restriction defines the
transition functions
\begin{equation}
\label{tfw}
t_{ij}: s_{ij} \to G
\end{equation}
that satisfy conditions analogous to \Ref{singlo}, \Ref{duppo},
\Ref{coci2}.  Denote $p_a$ and $p_b$ the end points of the segment
$s_{ij}$.  These are dual to the faces $a$ and $b$ of $\Delta^*$.
Partition each segment $s_{ij}$ into two parts, by choosing an
arbitrary point $p_{(ij)}$ on it.  Call $e_{a(ij)}$ the oriented
segment going from $a$ to $p_{(ij)}$.  Thus the segment $s_{ij}$ is
the composition of $e_{a(ij)}$ and $e^{-1}_{b(ij)}$. See Figure 2.

\begin{figure}
  \begin{center}
  \vskip.3cm
  \includegraphics[height=5cm]{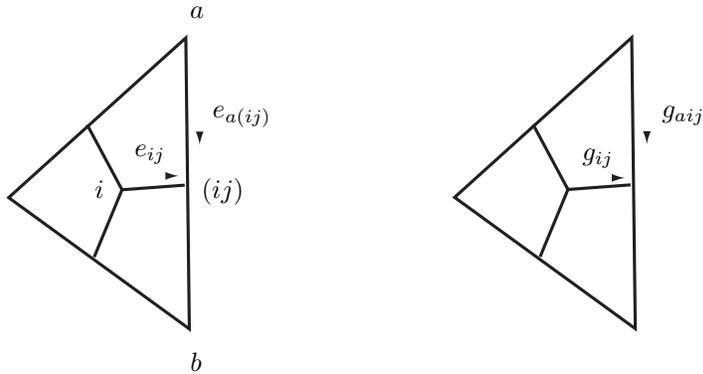}
  \end{center}
  \caption{\label{fig2} A triangle $i$ of the triangulation $\Delta$,
  divided in three wedges. The left panel indicates the names of the
  segments. The right panel indicates the group elements associated
  to these segments.}
  \end{figure}

A gauge transformation is a collection of maps $\lambda_{i}$ from the
perimeter of each triangle $i$ to $G$, such that its image of the
perimeter is contractible (because it must be possible to extend the
map continuously to the entire triangle).  It acts on the transition
functions in the following way,
\begin{equation}
\label{gaugetranszz}
t_{ij}(s) \longmapsto \lambda^{-1}_i(s)\
t_{ij}(s)\ \lambda_j(s).
\end{equation}
A principal bundle is defined by the transition functions (\ref{tfw})
up to these gauge transformations.

As before, the key quantity of interest is the {\em variation} of the
transition function along each half-segment.  Let us therefore define
the quantity
\begin{equation}
\label{gaugetransnonab2}
n_{a(ij)}= \frac{1}{2\pi} \int_{e_{a(ij)}} ds\ \ t^{-1}_{ij}(s)
\ \frac{d}{ds} t_{ij}(s)
\end{equation}
associated to each edge.  Restricting to the $G=SO(2)$ case we are
interested in, this can be expressed as
\begin{equation}
\label{gaugetrans222}
n_{a(ij)}= \frac{1}{2\pi}
\int_{e_{a(ij)}} d\psi_{a(ij)}(s),
\end{equation}
where $t_{a(ij)}(s)=e^{\psi_{a(ij)}(s)\tau}$.
We also define
\begin{equation}
\label{gaugetrans2222}
n_{ab}=n_{a(ij)}+n_{b(ij)}.
\end{equation}

A gauge transformation in the open set $i$ can add an arbitrary real
number $\lambda_{a(ij)}$ to each $n_{a(ij)}$.  This number is the
variation of the gauge transformation along $e_{a(ij)}$.  The gauge
transformation can be written as
\begin{equation}
\label{gaugetransd2}
n_{\alpha(ij)} \mapsto n_{\alpha(ij)} + \lambda^a_{i} -
\lambda^{j}_{i}
- \lambda^a_{j} + \lambda^i_{j}.
\end{equation}

As before, this structure can be simplified using a partial gauge
fixing, which makes it much easier to incorporate the cocycle
condition directly into the quantities $n_{a(ij)}$.  Because of the
cocycle condition (\ref{coci2}) it is always possible to chose a gauge
transformation \Ref{gaugetranszz} such that the transformed
$t_{ij}(s)$ is the identity at the vertices, as well as on the points
$p_{(ij)}$.  Denote this gauge the ``edge gauge".  Then the cocycle
condition is trivially satisfied.  In this gauge the transition
functions define maps $t_{\alpha(ij)}: S_{1}\to G$ and
\Ref{gaugetrans222} is the number of times the transition function of
the segment wraps around the group, and is an integer.  Within this
gauge, we can still gauge transform the integers $n_{a(ij)}$ by means
of a gauge transformation that is the identity on the vertices, but
the quantities $\lambda^a_{i}$ and $\lambda^j_{i}$ and in
\Ref{gaugetransd2} must be integers.

The bundle is uniquely characterized by the discrete set of {\em
integers} numbers $n_{a(ij)}$, and the gauge transformations
(\ref{gaugetransd2}).  The only invariant quantity is
\begin{equation}
\label{kdis2}
e(P) =\frac{1}{2}\sum_{a(ij)} n_{\alpha(ij)},
\end{equation}
which can be recognized again as the Euler number of the principal
bundle.

In order to write the discretization of the field theory, we use the
``derived complex'' \cite{Mau}.  Its utility for the spinfoam
formalism was recognized by Reisenberger in \cite{Mike2}.  This is
obtained by considering both a triangulation $\Delta$ of the manifold
and its dual cellular complex $\Delta^*$.  Denote $v_{i}=v_{(abc)}$ a
vertex of the dual complex $\Delta^*$ (an arbitrary point inside the
triangle $i$.)  Let $e_{ij}$ be a segment joining the point $v_{i}$
with the point $p_{(ij)}$.  Notice that the edges $e_{ab}$ of the dual
complex $\Delta^*$ are formed by the conjunctions of $e_{ij}$ and
$e^{-1}_{ji}$.  The intersection of the triangle $i$ (of $\Delta$) and
the face $a$ (of $\Delta^*$) is a quadrangle called a {\em wedge} and
denoted $w_{ai}$.  It is bounded by the four segments $e_{ij}$,
$e^{-1}_{a(ij)}$, $e_{a(ik)}$ and $e_{ik}$.  See Figure 2 and Figure
3.

We now introduce a discretization of the connection field $\omega$.
This will be given by two kinds of group elements.  First, the group
elements $g_{ij}$ associated with the oriented segments $e_{ij}$
\begin{equation}
g_{ij} = e^{\frac{k}{\hbar}\int_{e_{ij}}\omega_{i}},
\label{scaledcab22}
\end{equation}
where $\omega_{i}$ is the one-form representing the connection in the
open set to which the triangle $i$ belongs.  Second, the two group
elements $g_{aij}$ and $g_{aji}$, both associated with the segment
$e_{a(ij)}$, and interpreted as the exponential of the scaled
connection in the two open sets $i$ and $j$ respectively.
\begin{equation}
g_{aij} = e^{\frac{k}{\hbar}\int_{s_{a(ij)}}\omega_{i}}, \hspace{2em}
g_{aji} = e^{\frac{k}{\hbar}\int_{s_{a(ij)}}\omega_{j}}.
\label{scaledcab2}
\end{equation}
These two group elements are related by the transition functions
\begin{equation}
g_{aij}\ (g_{aji})^{-1} = e^{\frac{k}{\hbar}2\pi n_{a(ij)}\tau}.
\label{gabgba2}
\end{equation}

\begin{figure}
  \begin{center}
  \includegraphics[height=4cm]{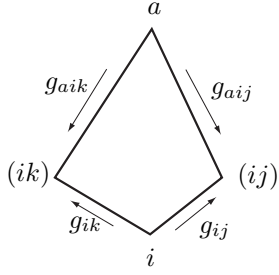}
  \end{center}
  \caption{\label{fig3} A wedge $w_{ai}$ and the group elements on
its
  perimeter.}
  \end{figure}

The discretized action can be written as sum over the wedges
$w_{ai}$.
We define the (exponentiated) curvature associated to each wedge $a$
as the product of the group elements associated to the segments
forming its perimeter.  See Figure \ref{fig3}.
\begin{equation}
\label{fprodlinksw}
g_{ai}=g^{\mbox{\ }}_{aij}\ g_{ij}^{-1}\ g^{\mbox{\ }}_{ik}\
g_{aki}^{-1}.
\end{equation}
The dynamics of the discrete model is defined by discretizing the
action (\ref{BF}) as
\begin{equation}
\label{actionw}
S_{BF}[B_{a},g_{ij},g_{aij}]=- \hbar\ \sum_{ai}\ {\rm
tr}\, [B^{\mbox{\ }}_{a}\ g^{\scriptscriptstyle{1/2}}_{ai}].
\end{equation}
This defines the discretized model in the second version.  Notice that
this sum can be entirely rewritten in terms of the group elements
\begin{equation}
g_{ab} = g_{ij}^{\mbox{\ }}\ g_{aij}^{-1}\ g_{aji}^{\mbox{\ }}\
g_{ji}^{-1},
\label{scalcon}
\end{equation}
which can be identified as the group elements denoted in the same
manner in section \ref{db}.  The relation between $g_{ab}$
and $g_{ba}$ follows from \Ref{gabgba2} and turns out to be precisely
given by \Ref{gabgba}, where $n_{ab}$ is now given by
\Ref{gaugetrans2222}.  This shows the equivalence with the previous
version.

The group elements $g_{ij}$ and $g_{aij}$ can be included in the GFT
formulation as follows (see \cite{mikecarlogft}.)  A 2d GFT
\cite{carlo}
is defined by a field $\phi(g_{1},g_{2})$ on $SO(2)\times SO(2)$, and
by the action
\begin{equation}
\label{actionGFT}
S[\phi] = \int dg_{1} dg_{2} \phi(g_{1},g_{2})\phi(g_{1},g_{2})
+\lambda \int dg_{1} dg_{2} dg_{3}dgdg'dg'' \phi(g_{1}g,g_{2}g)
\phi(g_{2}g',g_{3}g') \phi(g_{3}g'',g_{1}g'')
\end{equation}
(for convenience we have inserted the projection implementing
gauge invariance in the interaction term instead than in the kinetic
term.)  The Feynman expansion can be written in terms of the
propagator
\begin{equation}
K(g_{1},g_{2};g'_{1},g'_{2})=  \delta(g_{1},g'_{1})\
\delta(g_{2},g'_{2}),
\end{equation}
and the vertex
\begin{equation}
V(g_{1},g_{2};g'_{1},g'_{2};g''_{1},g''_{2})=\lambda\int dgdg'dg''
\delta(g_{1}g,g'_{1}g'')\ \delta(g_{2}g,g'_{2}g')
\delta(g_{3}g',g'_{3}g'').
\end{equation}
The Feynman combinatorics gives a sum over two-complexes because
sequences of delta functions with corresponding arguments define loops
on the Feynman graph that can be identified with faces \cite{GFT}.
Let us call $i$ the vertices and $a$ the faces of one two-complex.
Then the group element appearing in the propagator connecting the
vertex $i$ with the vertex $j$ and belonging to the face $a$ can be
called $g_{aij}$.  The group elements appearing in the integrals in
the vertex and belonging to end of the vertex $i$ that gets connected
with the vertex $j$ can be called $g_{ij}$.  It is then a simple
exercise to see that the weight of a two-complex is given by the
partition function of the spinfoam model considered above.

To take into account the possibility of a nontrivial bundle, in the
GFT formalism we can modify the kinetic term.  In fact, imagine to
replace the kinetic term in \Ref{actionGFT} as follows
\begin{equation}
\label{actionGFT2}
S[\phi] = \int dg_{1} dg_{2} \phi(g_{1}h,g_{2}h')\phi(g_{1},g_{2})
+\lambda \int dg_{1} dg_{2} dg_{3}dgdg'dg'' \phi(g_{1}g,g_{2}g)
\phi(g_{2}g',g_{3}g') \phi(g_{3}g'',g_{1}g'').
\end{equation}
where
\begin{equation}
h = e^{\frac{k}{\hbar}2\pi n\tau},\hspace{1cm} h' =
e^{\frac{k}{\hbar}2\pi n'\tau}.
\end{equation}
Then the propagator becomes
\begin{equation}
K(g_{1},g_{2};g'_{1},g'_{2})=  \delta(g_{1}h,g'_{1})\
\delta(g_{2}h',g'_{2}).
\end{equation}
This modifies the spinfoam model simply by replacing
\begin{equation}
\delta(g_{aij},g_{aji})
\end{equation}
with
\begin{equation}
\delta(g_{aij}h,g_{aji}).
\end{equation}
That is, it implements \Ref{gabgba2}, where $h$ can be identified with
$e^{\frac{k}{\hbar}2\pi n_{a(ij)}\tau}$.  However, such a modification
would give all transition function equal to one another.  To have
different transition functions on different segments we have to
consider
the transition functions as dynamical variables and sum over them.
Thus, introducing the bundle topology in the GFT appears to require a
sum over the topologies of the fiber bundles.  This can be achieved by
considering the action
\begin{eqnarray}
\label{actionGFT22}
S[\phi] &=& \sum_{n,n'} c_{n}c_{n'}\int dg_{1} dg_{2}\ \
\phi(g_{1}e^{\frac{k}{\hbar}2\pi n\tau},g_{2}e^{\frac{k}{\hbar}2\pi
n'\tau})\ \phi(g_{1},g_{2}) \nonumber\\
&&\hspace{1cm}+\ \lambda \int dg_{1} dg_{2} dg_{3}dgdg'dg''\
\phi(g_{1}g,g_{2}g)\ \phi(g_{2}g',g_{3}g')\ \phi(g_{3}g'',g_{1}g''),
\end{eqnarray}
for suitable weights $c_{n}$.  The Feynman expansion of this action
gives a spinfoam model that involves a sum over (topologically
inequivalent) two-complexes as well as over all fiber
bundle topologies.  The properties of this sum will be studied
elsewhere.

\end{document}